\documentclass[prl,twocolumn,showpacs,superscriptaddress]{revtex4-1}

\newcommand{\Equation}[1]{Equation~(\ref{eq:#1})}
\newcommand{\Eq}[1]{Eq.~(\ref{eq:#1})}

\newcommand{\Fig}[1]{Fig.~\ref{fig:#1}}

\newcommand{\Ref}[1]{Ref.~\cite{#1}}

\newcommand{\Figure}[1]{Figure~\ref{fig:#1}}
\newcommand{\rr}{\mathbf{r}}

\renewcommand{\O}{{\cal O}}
\newcommand{\gdot}{\dot{\gamma}}
\newcommand{\qO}{{q_\O}}
\newcommand{\qzero}{q_1}
\newcommand{\gzero}{g_1}
\newcommand{\phiJzero}{\phi_{J1}}
\newcommand{\qsigma}{{q_\sigma}}
\usepackage{graphicx}

\begin{document}
\title{Critical Scaling of Shearing Rheology at the Jamming Transition of Soft
  Core Frictionless Disks}

\author{Peter Olsson}

\affiliation{Department of Physics, Ume\aa\ University, 
  901 87 Ume\aa, Sweden}

\author{S. Teitel}

\affiliation{Department of Physics and Astronomy, University of Rochester,
  Rochester, NY 14627}

\date{\today}   

\begin{abstract}
  We perform numerical simulations to determine the shear stress and pressure of
  steady-state shear flow in a soft-disk model in two dimensions at zero
  temperature in the vicinity of the jamming transition $\phi_J$. We use
  critical point scaling analyses to determine the critical behavior at jamming,
  and we find that it is crucial to include {\it corrections to scaling} for a
  reliable analysis.  We find that the relative size of these corrections are
  much smaller for pressure than for shear stress. We furthermore find a
  superlinear behavior for pressure and shear stress above $\phi_J$, both from
  the scaling analysis and from a direct analysis of pressure data extrapolated
  to the limit of vanishing shear rate.
\end{abstract}

\maketitle

Granular materials, supercooled liquids, and foams are examples of systems that may
undergo a transition from a liquid-like to an amorphous solid state as some
control parameter is varied. It has been hypothesised that the transitions in
these strikingly different systems are controlled by the same mechanism
\cite{Liu_Nagel} and the term jamming has been coined for this transition.

Much work on jamming has focused on a particularly simple model, consisting of
frictionless spherical particles with repulsive contact interactions at zero
temperature \cite{OHern_Silbert_Liu_Nagel:2003}. The packing fraction (density)
of particles $\phi$ is the key control parameter in such systems.  Jamming upon
compression, and jamming by relaxation from initially random states, have been
the focus of many investigations \cite{OHern_Silbert_Liu_Nagel:2003,
  Chaudhuri_Berthier_Sastry,Vagberg_VMOT:qs-scaling}.  Another, physically
realizable and important case is jamming upon shear deformation.  This has been
modeled both by simulations at a finite constant shear strain rate $\gdot$
\cite{Olsson_Teitel:jamming,Hatano:2008, Hatano:2008:arXiv,
  Otsuki_Hayakawa:2009b, Hatano:2009, Hatano:2010, Tighe_WRvSvH}, as well as by
quasistatic shearing \cite{Vagberg_VMOT:qs-scaling, Heussinger_Barrat:2009,
  Heussinger_Chaudhuri_Barrat-Softmatter}, in which the system relaxes to its
energy minimum after each finite small strain increment.

Several attempts have been made to determine the critical packing fraction $\phi_J$
and critical exponents, describing behavior at shear driven jamming
\cite{Olsson_Teitel:jamming, Hatano:2008, Hatano:2008:arXiv,
  Otsuki_Hayakawa:2009b, Hatano:2010, Tighe_WRvSvH}. There is however little
agreement on the values of the exponents and there is thus a need for a thorough
and careful investigation of the jamming transition in the shearing ensemble.
It will also be interesting to compare the exponents found from shearing
rheology to those found from compressing marginally jammed packings.  In
particular we note the linear increase of pressure above jamming that is
observed in that system \cite{OHern_Silbert_Liu_Nagel:2003,
  Chaudhuri_Berthier_Sastry}, compared to the superlinear behavior often
reported in the sheared system for pressure and/or shear stress
\cite{Olsson_Teitel:jamming, Hatano:2008,  Tighe_WRvSvH,
  Hatano:2010}.

In this Letter we do a careful scaling analysis of high precision data for both
shear stress and pressure at shear strain rates down to $\gdot=10^{-8}$. Instead of
relying on visually acceptable data collapses we use a non-linear minimization
technique to determine the best fitting parameters.  As in a recent analysis of
energy-minimized configurations \cite{Vagberg_VMOT:qs-scaling} we find that it
is necessary to include {\it corrections to scaling}, but also that the magnitude of
the corrections are markedly different for different quantities, and,
furthermore, that the neglect of these corrections is a major reason for the
differing values for the critical exponents in the literature.  We find strong
evidence for a superlinear behavior of yield stress and pressure above jamming
from the scaling analysis, and also find independent support for this result
from pressure data extrapolated to the limit of vanishing shear rate. We also
suggest a possible mechanism behind this behavior.

Following O'Hern \emph{et al.}\ \cite{OHern_Silbert_Liu_Nagel:2003} we use a
simple model of bi-disperse frictionless soft disks in two dimensions 
with equal numbers of disks with two different radii in the ratio
1.4. Length is measured in units of the diameter of the small particles, $d_s$. With
$r_{ij}$ the distance between the centers of two particles and $d_{ij}$ the
sum of their radii, the interaction between overlapping particles is $V(r_{ij})
= (\epsilon/2) (1 - r_{ij}/d_{ij})^2$.  We use Lees-Edwards boundary conditions
\cite{Evans_Morriss} to introduce a time-dependent shear strain $\gamma =
t\gdot$. With periodic boundary conditions on the coordinates $x_i$ and $y_i$ in
an $L\times L$ system, the position of particle $i$ in a box with strain
$\gamma$ is defined as $\rr_i = (x_i+\gamma y_i, y_i)$.  We simulate overdamped
dynamics at zero temperature with the equation of motion \cite{Durian:1995},
\begin{displaymath}
  \frac{d\rr_i}{dt} = -{C}\sum_j\frac{dV(\rr_{ij})}{d\rr_i} + y_i \gdot\; \hat{x},
\end{displaymath}
with $\epsilon=1$ and $C=1$.  The unit of time is $\tau_0 = d_s/{C}\epsilon$.
All our simulations at the lower shear rates are from $N\geq65536$ total particles.

Our basic scaling assumption describes how different quantities, as e.g.\ shear
stress, pressure, potential energy and jamming fraction, depend on a change of
length scale with a scale factor $b$:
\begin{equation}
  \O(\delta\phi, \gdot, 1/L) = b^{-y_\O/\nu} g_\O(\delta\phi b^{1/\nu},
  \gdot b^z, b/L).
  \label{eq:O-scale}
\end{equation}
Here $\delta\phi=\phi-\phi_J$, $y_\O$ is the critical exponent of the observable
$\O$, $\nu$ is the correlation length exponent, and $z$ is the dynamic critical
exponent.  Point J is at $\delta\phi=0$, $\gdot\rightarrow0$, and an infinite
system size, $1/L\rightarrow0$; the scaling relation describes the departure
from the critical point in these respective directions.

The above expression may be used as a starting point for our analysis.  We make
use of data obtained at finite shear rates and system sizes large enough that
finite size effects may be neglected --- essentially the same approach as in
Ref.~\cite{Olsson_Teitel:jamming}. With $b=\gdot^{-1/z}$ in \Eq{O-scale} and
$q_\O \equiv y_\O/z\nu$, the scaling relation becomes
\begin{equation}
  \O(\delta\phi, \gdot) \sim \gdot^{\qO} g_\O(\delta\phi/ \gdot^{1/z\nu}),
  \label{eq:O-gdot-scale}
\end{equation}
where the scaling function is a function of only a single argument.
At $\phi_J$ we have $\O(\phi_J,\gdot)\sim\gdot^\qO$ which gives a simple method
for determining $\qO$ and $\phi_J$: Plot $\O$ versus $\gdot$ on a double-log
scale for several different $\phi$.  The packing fraction for which the data
fall on a straight line is then our estimated $\phi_J$. Data above and below
$\phi_J$, respectively, should curve in opposite directions.

\begin{figure}
  \includegraphics[width=7cm]{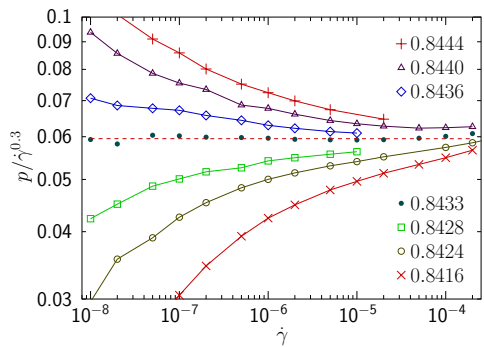}
  \caption{Approximate determination of $\phi_J$ and $q_p$ from
    \Eq{O-gdot-scale} without corrections to scaling.  The figure shows pressure
    versus shear rate at several different packing fractions.  The pressure is
    shown as $p/\gdot^{0.3}$ to make the behavior clearly visible. This suggests
    that $p\sim\gdot^{\qzero}$ with $\qzero= 0.30$ at $\phiJzero= 0.8433$.}
  \label{fig:p-gdot}
\end{figure}

We start by applying this simple recipe to the pressure, $p$, and will turn to
the shear stress only as the next step. Both these quantities are calculated, as
in \Ref{OHern_Silbert_Liu_Nagel:2003}, from the elastic forces
only. \Figure{p-gdot} shows pressure versus shear rate at several different
packing fractions. Anticipating that the value of $q_p \approx 0.3$, we plot
$p/\gdot^{0.3}$ vs $\gdot$ in order to more clearly differentiate the behaviors
near $\phi_J$. It is then easy to identify the density
with a rectilinear behavior, and we find $p\sim\gdot^{\qzero}$ with $\qzero= 0.3$ at $\phiJzero= 0.8433$.
Data at lower and higher densities curve downwards and upwards,
respectively. These values $\phiJzero$ and $\qzero$ are only first
estimates of the jamming density and the exponent, respectively; our final
estimates turn out to be just slightly different.

\begin{figure}
  \includegraphics[width=7cm]{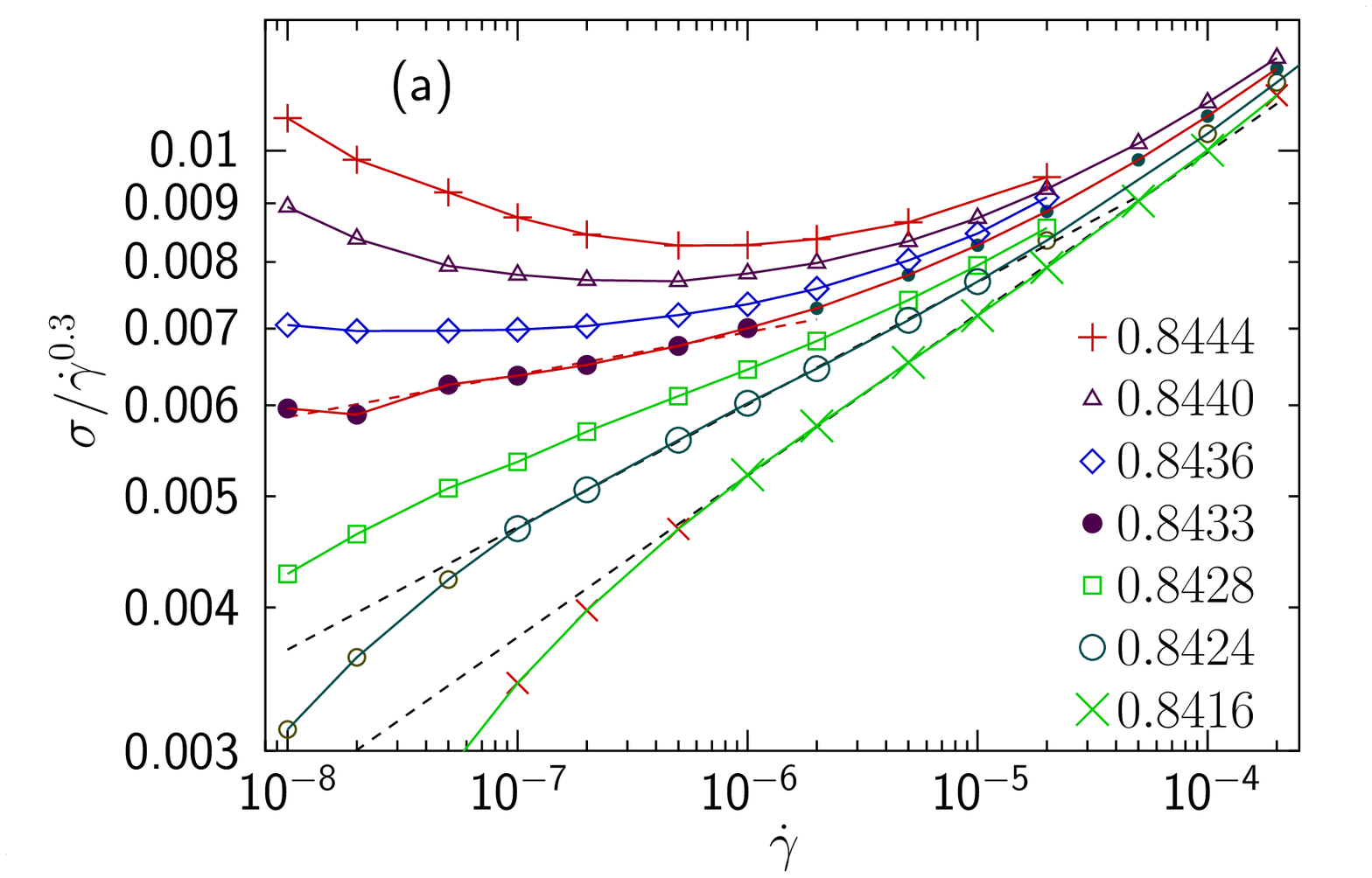}
  \includegraphics[width=7cm]{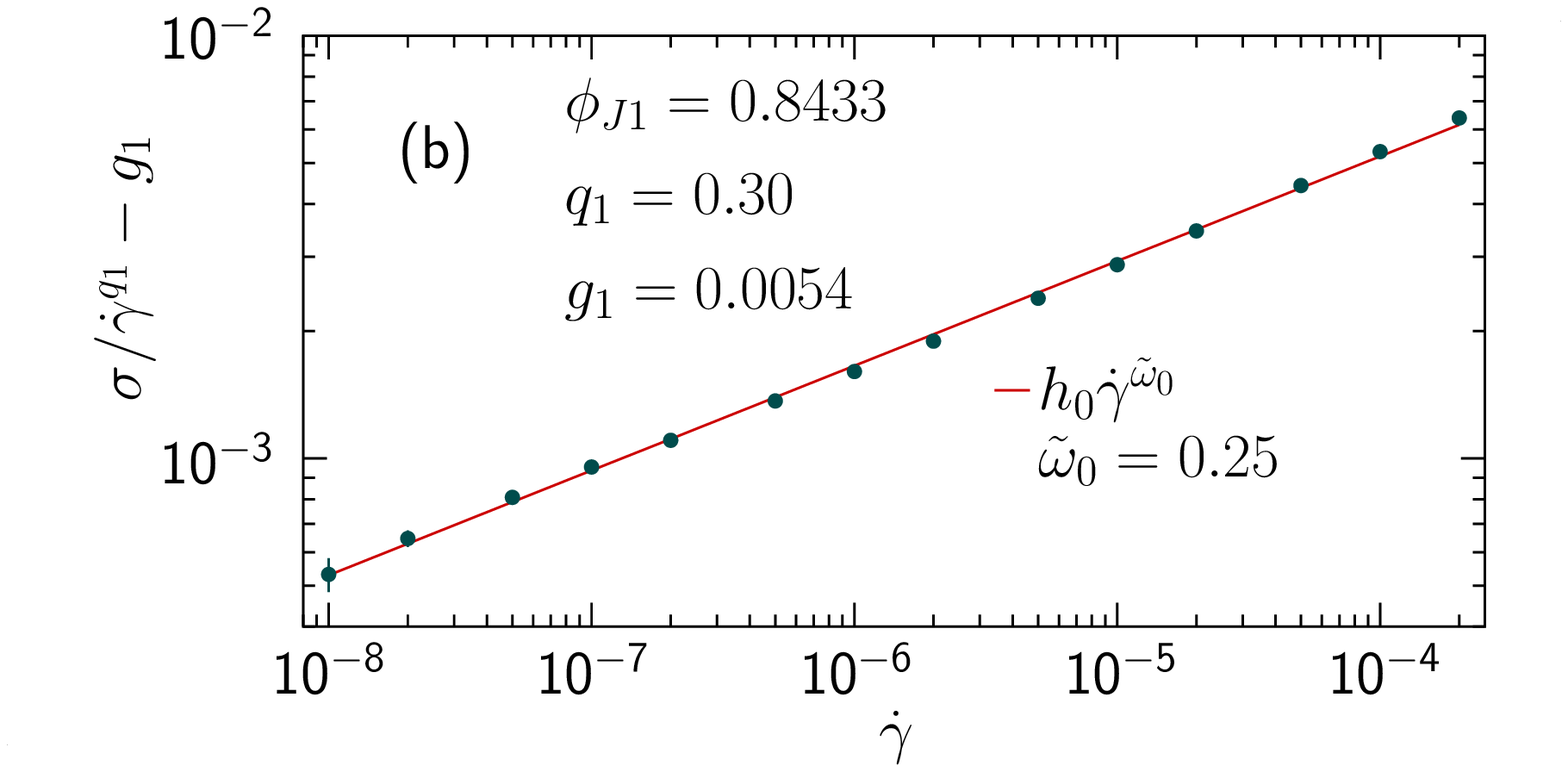}
  \caption{Shear stress $\sigma$ versus shear rate $\dot\gamma$ at several
    different densities.  Panel (a) shows that there is no density where
    $\sigma$ behaves algebraically across the extended range of shear rates,
    however data across two orders of magnitude of $\gdot$ could, to a
    reasonable approximation, be taken as algebraic. In that vein, data limited
    to $10^{-6}\le\dot\gamma$ gives $\phi=0.8416$ (crosses) as a good candidate
    for $\phi_J$ (cf.\ \Ref{Olsson_Teitel:jamming}) whereas other ranges of
    $\gdot$ would give other estimates.  From a comparison with
    $p/\gdot^{\qzero}=$ const at $\phiJzero$ in \Fig{p-gdot}, panel (b) shows
    the correction term $\sigma/\gdot^{\qzero} - \gzero$ at $\phiJzero$, and it
    appears that this correction to a very good approximation is
    $\sim\gdot^{\tilde\omega_0}$, which has the same form as standard
    corrections to scaling in critical phenomena.}
  \label{fig:sigma-gdot-first}
\end{figure}

\Figure{sigma-gdot-first} is the same kind of plot for the shear stress,
$\sigma$, and it is immediately clear that these data are not directly amenable to
the same kind of analysis; there is no density with an algebraic behavor across
the whole range of shear rates. Before presenting our further analyses we note
that this provides an explanation for the differing values of both jamming
density and exponents in the literature. In \Ref{Olsson_Teitel:jamming} the
jamming density was found to be $\approx 0.8415$ and the figure shows that data
in the range $10^{-6}\leq \gdot \leq10^{-4}$ would suggest $\phi=0.8416$
(crosses) as a good candidate for $\phi_J$. However, it is clear that data at
the same density and lower shear rates deviate from the algebraic
behavior. Similarly, with access to $\sigma$ down to $\gdot=10^{-7}$,
$\phi=0.8424$ (open circles) would appear as a good candidate for $\phi_J$,
whereas data in the range $10^{-8}\leq\gdot\leq 10^{-6}$ would suggest
$\phi_J=0.8433$ (solid dots). The value of the effective exponent $\qsigma$ also changes:
for these three different ranges of shear rates we find $\qsigma = 0.44$, 0.41,
and 0.33, respectively. Note that this explanation is at variance with
\Ref{Tighe_WRvSvH} where the differing exponents are attributed to using data
from a too large range in $\phi$. That explanation is not applicable here since
our analyses only consider data right at the presumed $\phi_J$.

As a step towards the final analysis we now consider the shear stress at
$\phiJzero$ and focus on the deviation from the algebraic behavior
$\sim\gdot^{\qzero}$. From \Fig{sigma-gdot-first}(a) we note that
$\sigma/\gdot^{\qzero}$ in the limit of low $\gdot$ appears to saturate at a
finite value, $0.005 < \gzero < 0.006$ and so we plot $\sigma/\gdot^{\qzero} -
\gzero$ in \Fig{sigma-gdot-first}(b). It is then possible to adjust $\gzero$ such that
the remainder is algebraic in $\gdot$,
\begin{equation}
  \sigma(\phi_{J,0}, \gdot)/\gdot^{\qzero} = \gzero + \gdot^{\tilde\omega_0} h_0,
  \label{eq:sigma-phiJ-wcorr}
\end{equation}
with the exponent $\tilde\omega_0\approx0.25$. 

The importance of this observation lies in the fact that standard {\it corrections to
scaling} modify Eq.~(\ref{eq:O-scale}) to give precisely this form \cite{Binder:1981}, 
\begin{displaymath}
  \label{eq:O-scale-corr}
  \O(\delta\phi, \gdot) / b^{y_\O/\nu} = g_\O(\delta\phi b^{1/\nu}, \gdot b^z)
    + b^{-\omega} h_\O(\delta\phi b^{1/\nu}, \gdot b^z),
\end{displaymath}
where $h_\O$ is another scaling function and $\omega$ is the correction to
scaling exponent.  Using $b=\gdot^{-1/z}$ in the above then gives
\begin{equation}
  \label{eq:O-wcorr}
  \O(\delta\phi, \gdot)/ \gdot^{q_\O} =
    g_\O(\delta\phi/ \gdot^{1/z\nu}) + \gdot^{\omega/z} h_\O(\delta\phi/\gdot^{1/z\nu}).
\end{equation}
\Equation{sigma-phiJ-wcorr} is just the special case when $\delta\phi=0$.

The above analysis of $\sigma$ relied on $\phiJzero$ and $\qzero$ determined
from the pressure data without corrections to scaling. We now set out to analyze
both pressure and shear stress directly from the scaling relation, \Eq{O-wcorr},
that includes the correction term, and determine the $\phi_J$, $q_\O$, $1/z\nu$,
and $\omega/z$ that allow for the best fit to \Eq{O-wcorr}. Here $g_\O$ and
$h_\O$ are scaling functions which we approximate with fifth-order polynomials
in $x \equiv \delta\phi/\gdot^{1/z\nu}$. The actual fits are done by minimizing
$\chi^2/\mathrm{dof}$ with a Levenberg-Marquardt method.  The number of points
in the fits range from about 100 to 250 depending on the range of data used in
the fits.

In this kind of involved analysis it is crucial to validate the results and to
that end we use several different criteria: (i) The first is to check the
quality of the fits: Are the deviations of the data from the scaling function
consistent with the statistical uncertainties?  We use $\chi^2/\mathrm{dof}$,
which should be close to unity to get a quantitative measure. (ii) A good
quality of the fit does however not by itself guarantee that the results are
reliable.  The second check is therefore whether the fitting parameters are
reasonably independent of the precise range of the data included in the fit. We
do this by systematically varying both the range of shear rates and the range of
densities; fixing $X=(\phi-0.8434)/\gdot^{0.26}$ we use the criterion
$|X|<X_\mathrm{max}$ with $X_\mathrm{max} = 0.2$, 0.3, and 0.4. This restriction
does not reflect the size of the critical region but rather that the
polynomial approximation of the scaling function breaks down for too large $X$.
(iii) A final check is whether the critical parameters from analyses of
different quantities (here $p$ and $\sigma$) agree with one another.

\begin{figure}
  \includegraphics[width=4.1cm]{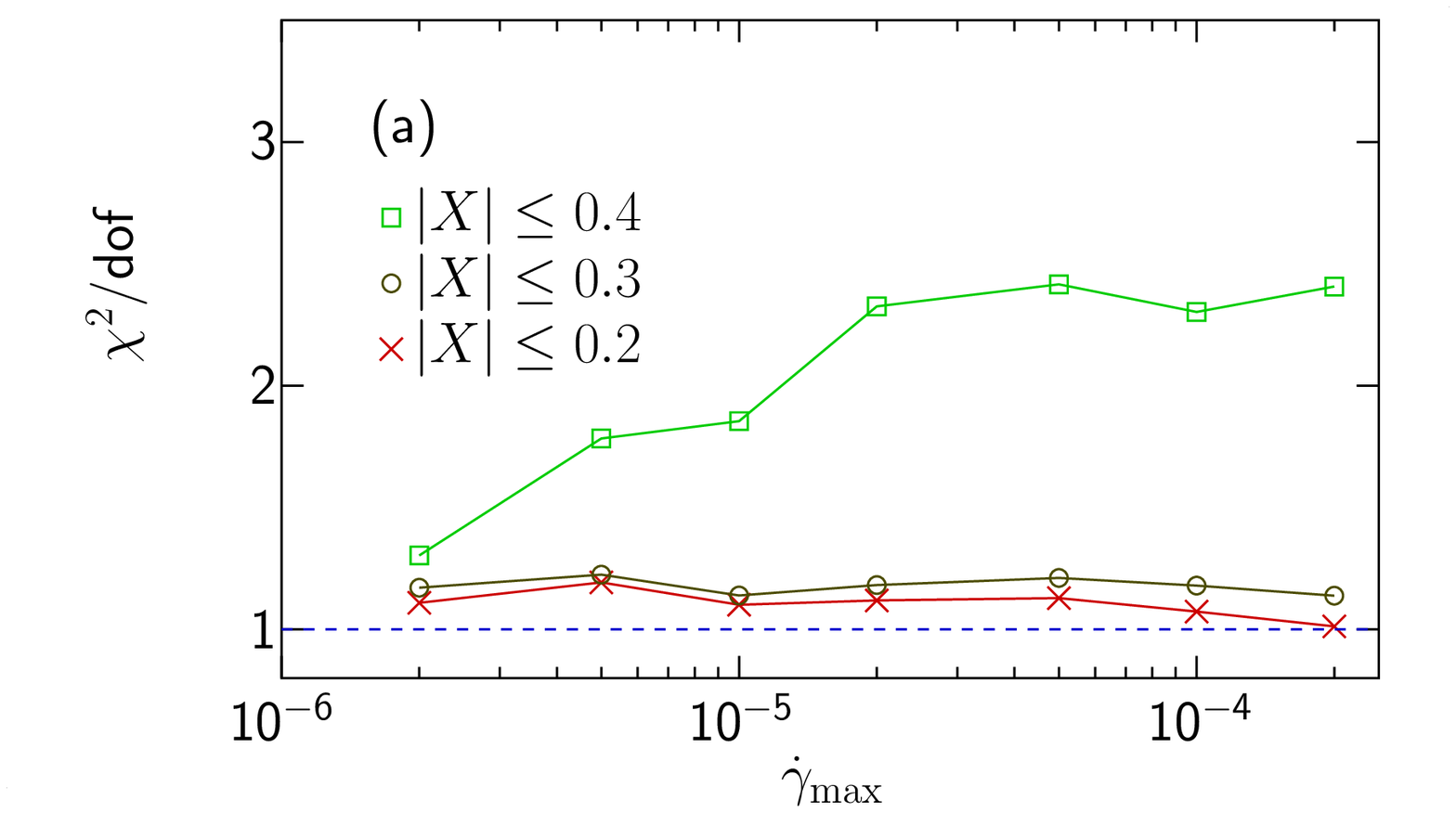}
  \includegraphics[width=4.1cm]{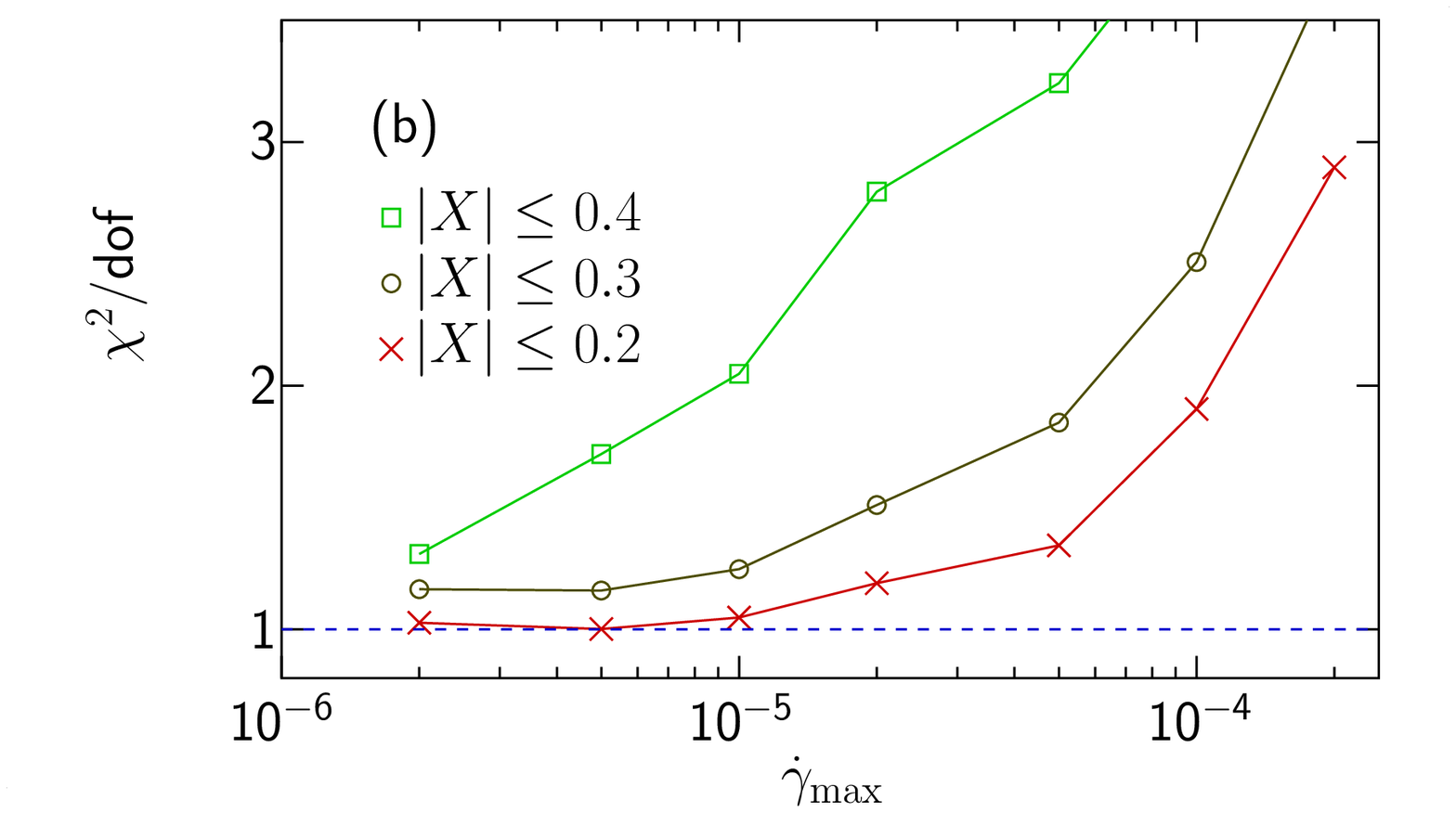}

  \includegraphics[width=4.1cm]{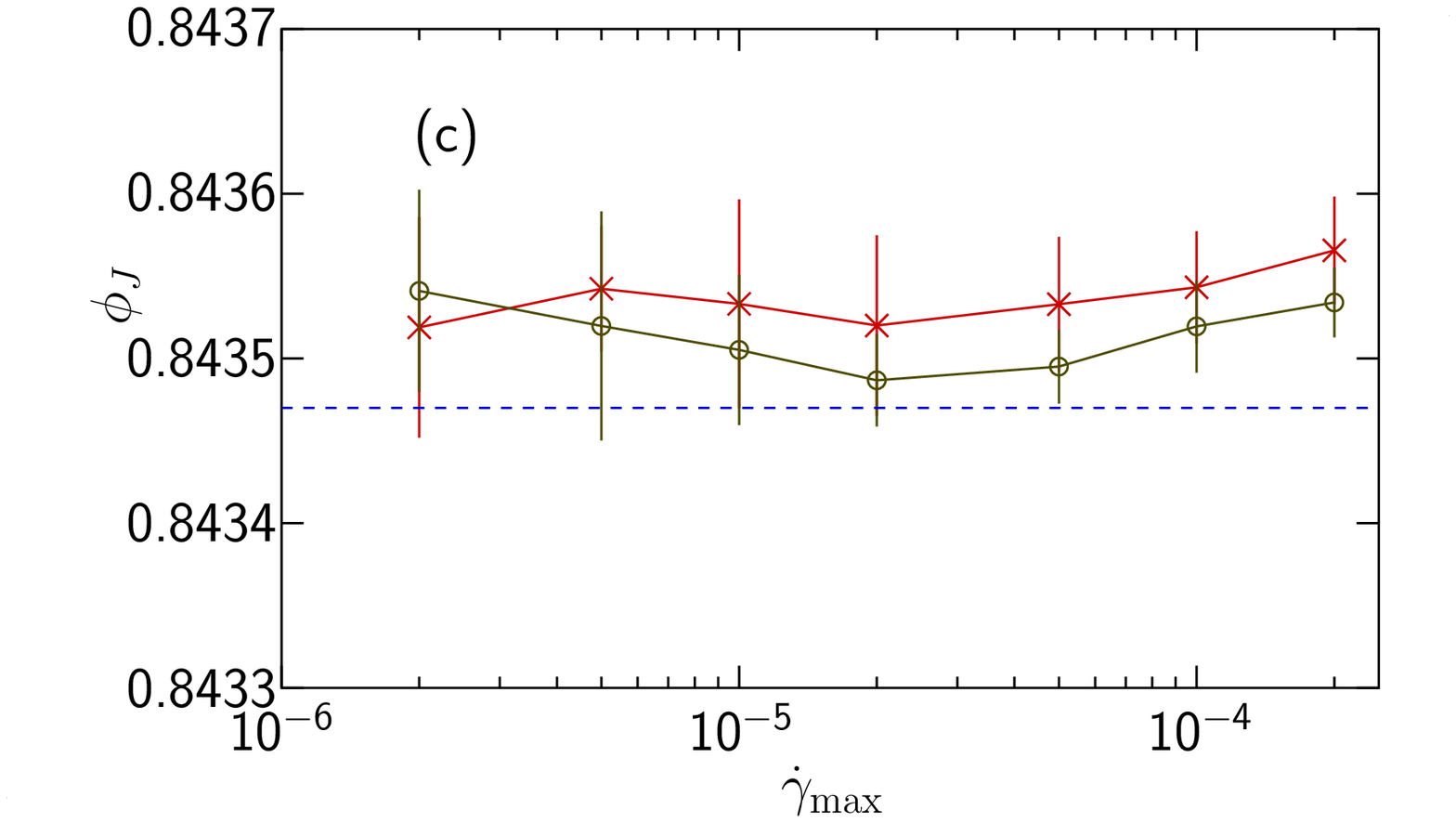}
  \includegraphics[width=4.1cm]{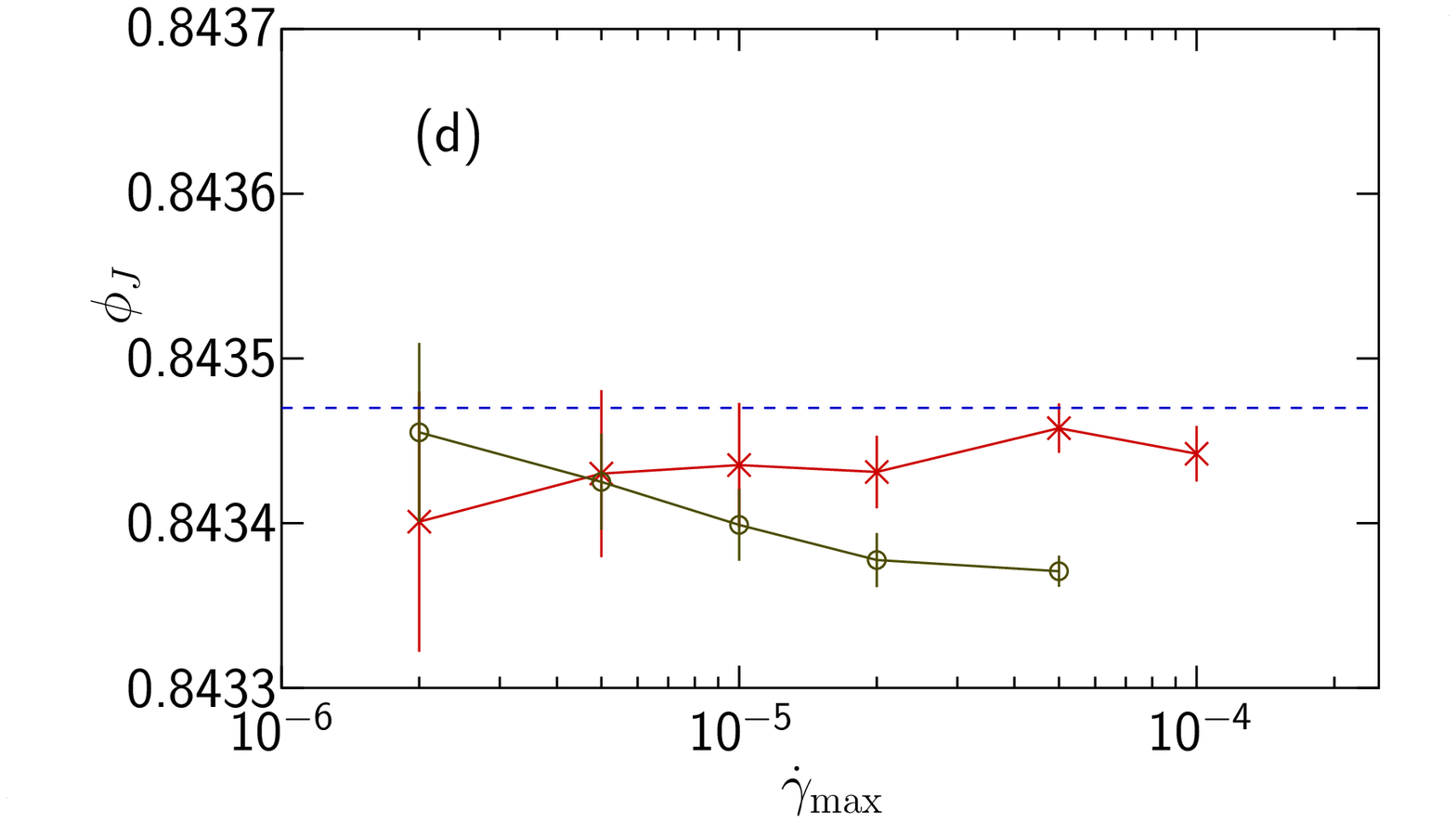}

  \includegraphics[width=4.1cm]{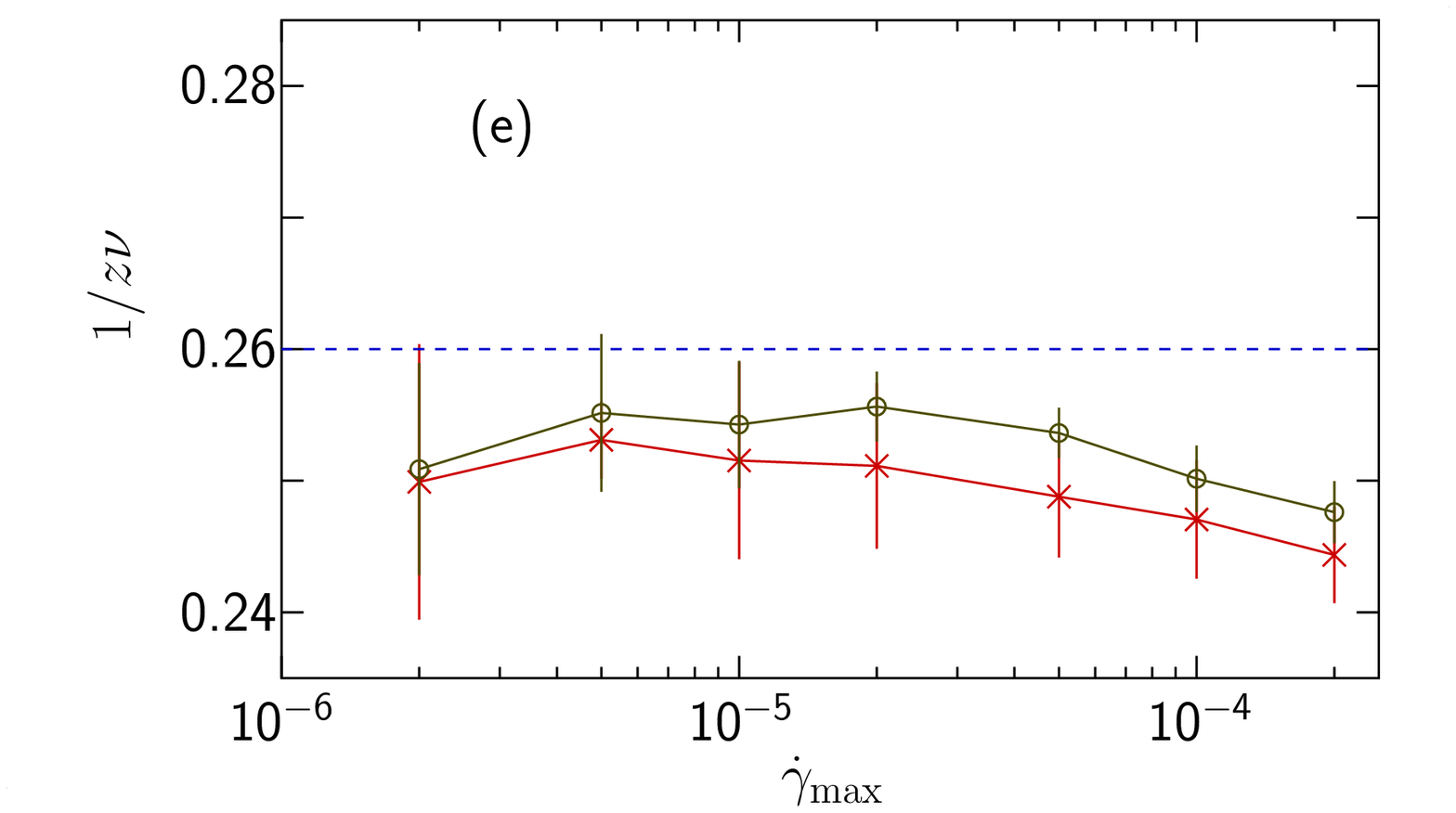}
  \includegraphics[width=4.1cm]{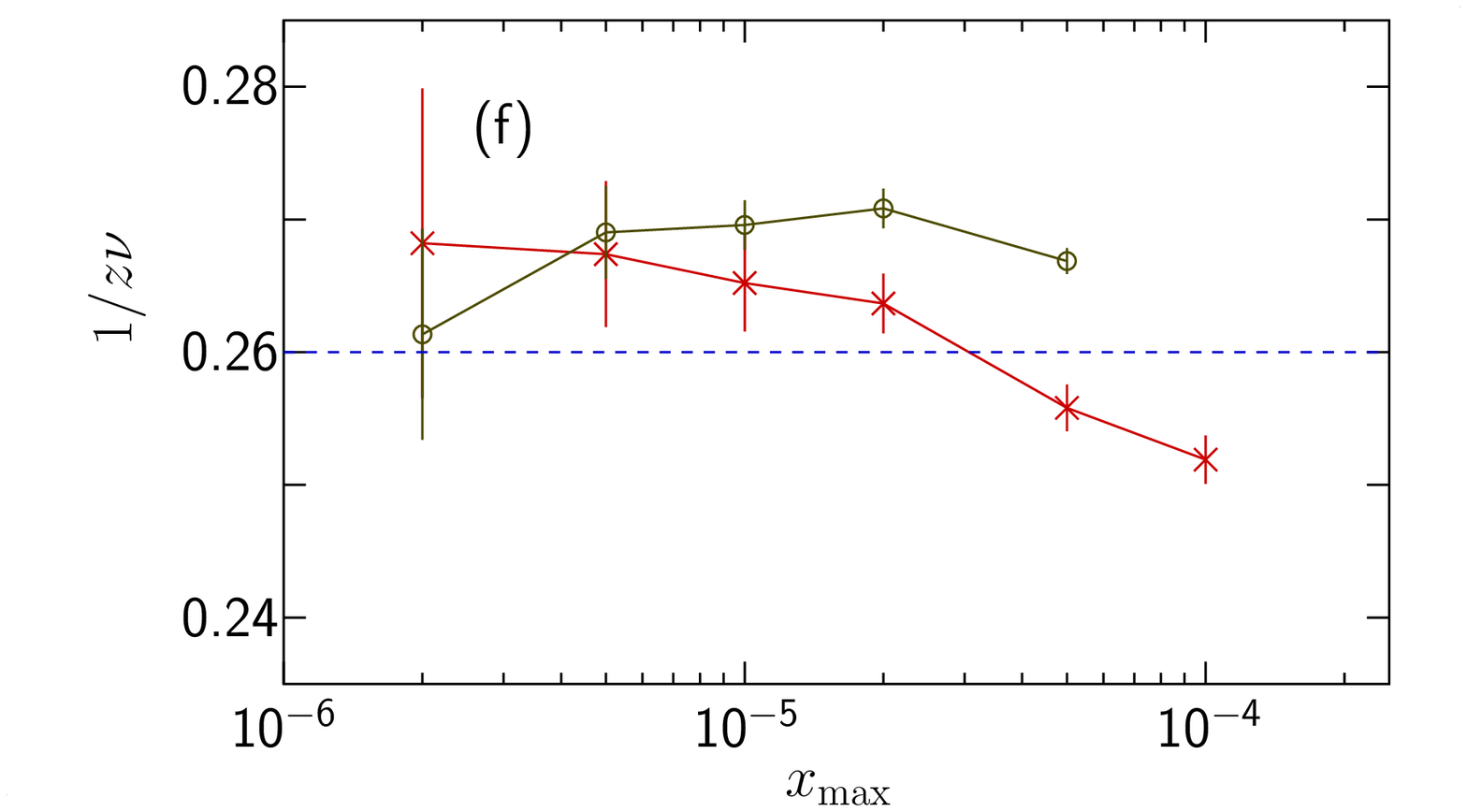}

  \includegraphics[width=4.1cm]{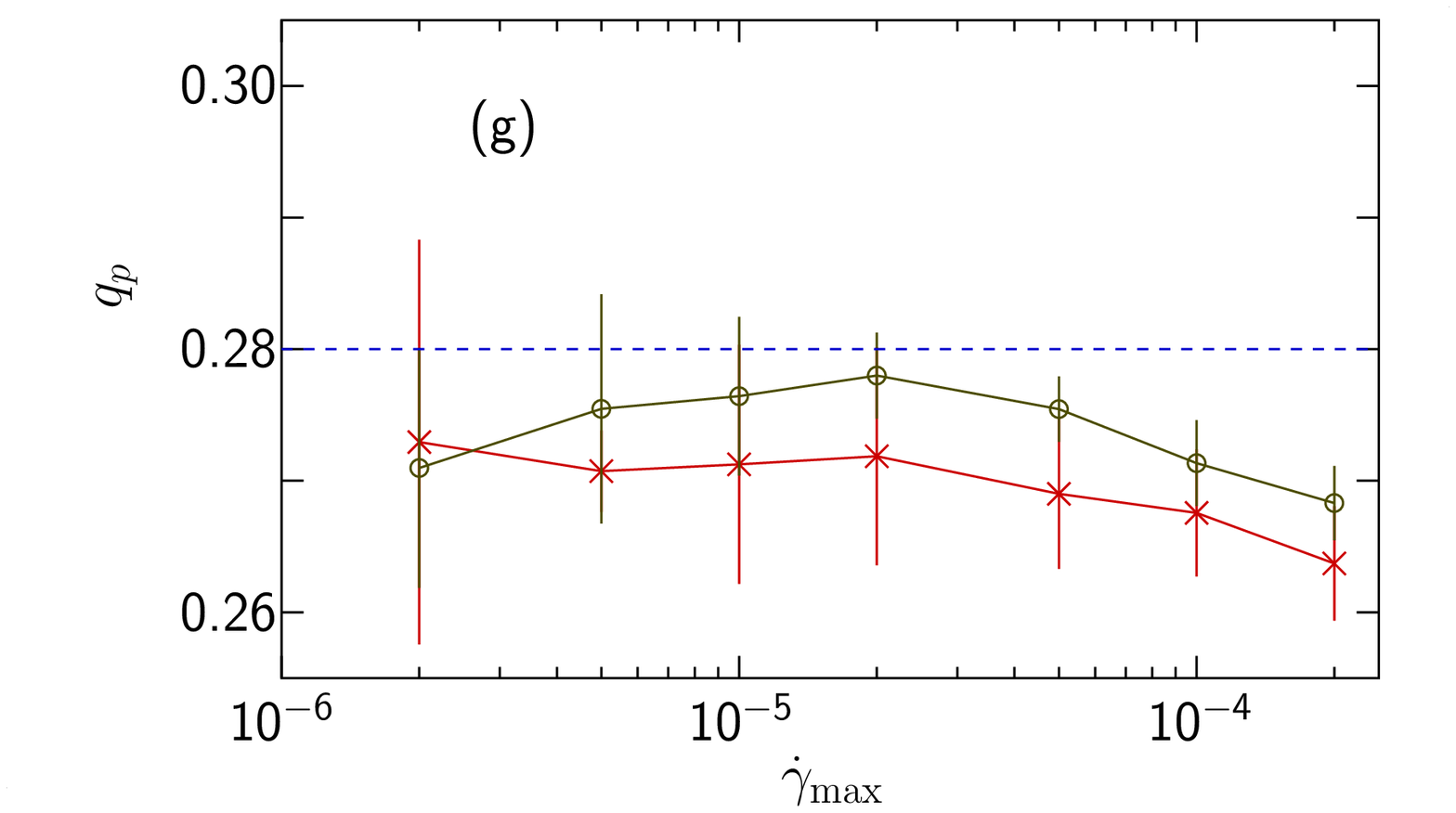}
  \includegraphics[width=4.1cm]{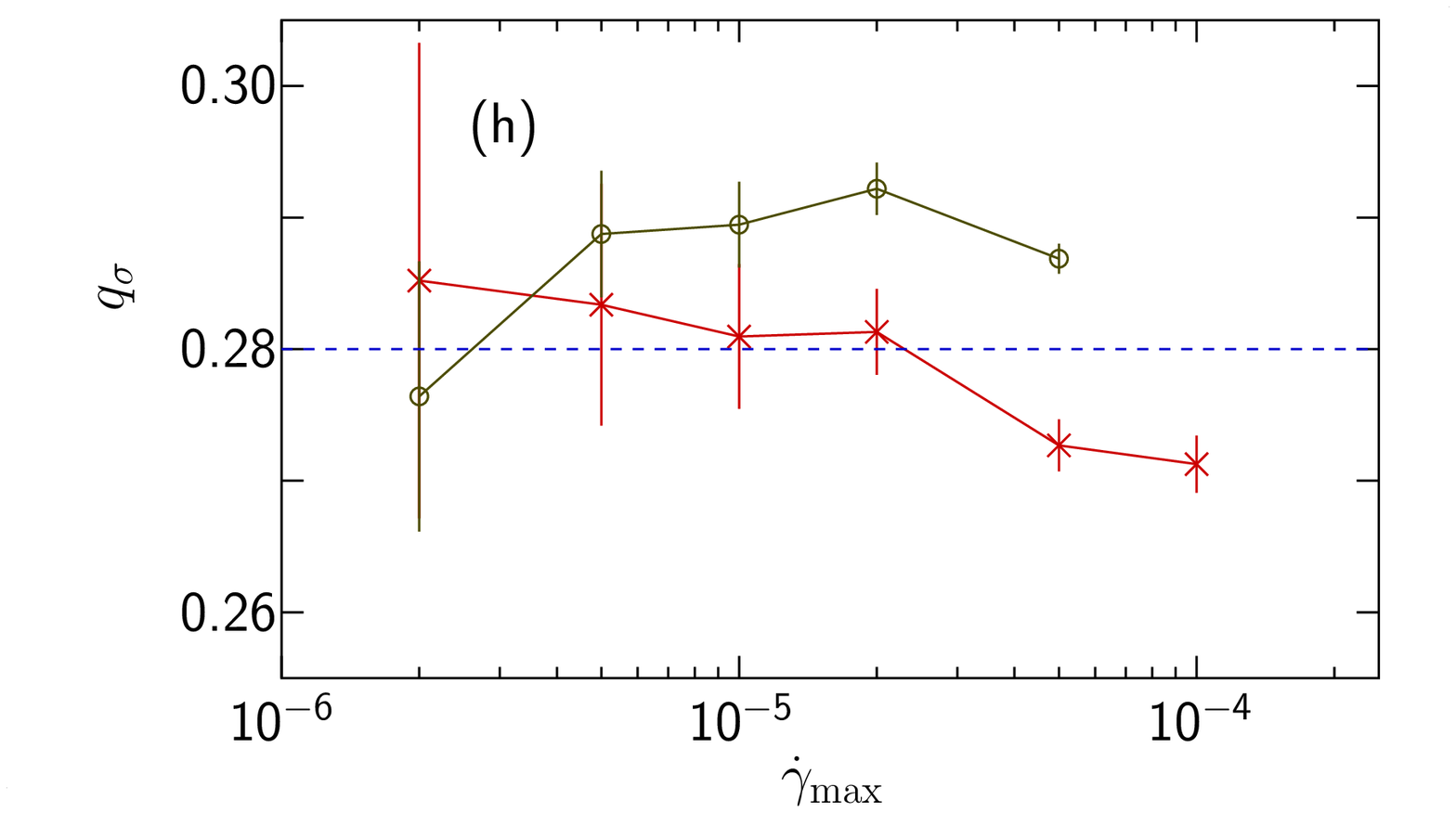}

  \caption{Results from scaling analyses that include corrections to
    scaling. The left and right panels are from analyses of pressure and shear
    stress, respectively. The first pair of panels, which give $\chi^2$/dof,
    suggest that the analyses are only reliable when data is used in a rather
    restrictive interval of $\phi-\phi_J$, $|X|\leq0.3$. For shear stress one
    also has to be restrictive in using data with larger $\gdot$. From the
    following panels we read off $\phi_J=0.84347$, $1/z\nu=0.26$, and
    $q_p=q_\sigma=0.28$. Combining the last two ($y=qz\nu$) gives $y_p=y_\sigma=1.08$.}
  \label{fig:coll-corr}
\end{figure}

Figures~\ref{fig:coll-corr} show $\chi^2/$dof and the key fitting parameters,
$\phi_J$, $1/z\nu$, $q_p$ and $q_\sigma$ plotted against
$\gdot_\mathrm{max}$. For each quantity the left and right panels are from
analyses of pressure and shear stress, respectively. First considering
$\chi^2/\mathrm{dof}$ in the first pair of panels, we note that the fits are
only good when the data are taken from a rather restrictive interval in $\phi$
around $\phi_J$, $|X|\leq0.3$. For pressure there is a good fit to the data over a
very large interval---more than four decades in $\gdot$. For the shear stress
the highest shear rates should not be used, and reliable results are obtained by
restricting $\gdot$ to $\gdot\leq5\times 10^{-5}$ when $X_\mathrm{max}=0.2$ and
$\gdot\leq1\times 10^{-5}$ for $X_\mathrm{max}=0.3$.

The next two panels show $\phi_J$ from pressure and shear stress, respectively,
in good agreement with one another; we estimate $\phi_J=0.84347\pm0.00020$ in
agreement with other recent determinations of $\phi_J$ from quasistatic
simulations\cite{Heussinger_Barrat:2009, Vagberg_VMOT:qs-scaling}. Here and
throughout, the error bars in the figures are one standard deviation whereas the
numerical values give a min--max interval ($\pm$ three standard deviations) for
the estimated quantities.  To correctly interpret these figures one should note
that the fitted values for different $\dot\gamma_{\rm max}$ and $X_{\rm max}$
are based on different subsets of the same data, and therefore are expected to
be strongly correlated. The main point is here to check how robust the fitting
parameters are to changes in the precise data set, and the absence of clear
trends in the results is therefore an encouraging sign.

We further find $1/z\nu=0.26\pm0.02$ and $q=0.28\pm0.02$.  Combining the two
exponents we find $y=qz\nu = 1.08\pm0.03$ (a strong correlation between $q$ and
$1/z\nu$ is responsible for the error estimate for $y$).
Since $y$ is just slightly above unity we have also reanalyzed the pressure data
with the assumption $y_p=1$. The fits then become considerably worse and we
conclude that the data is strongly in favor of $y_p>1$. A similar analysis of
the shear stress is not conclusive. Using $\nu=1.09$ from
\Ref{Vagberg_VMOT:qs-scaling} the dynamic critical exponent becomes
$z=3.5\pm0.4$. The correction to scaling exponent (not shown) is
$\omega/z=0.29\pm0.03$, or $\omega\nu=1.10\pm0.06$, which, again using
$\nu=1.09$ \cite{Vagberg_VMOT:qs-scaling}, gives $\omega=1.0\pm0.1$ in good
agreement with \Ref{Vagberg_VMOT:qs-scaling}.

\begin{figure}
  \includegraphics[width=4.1cm]{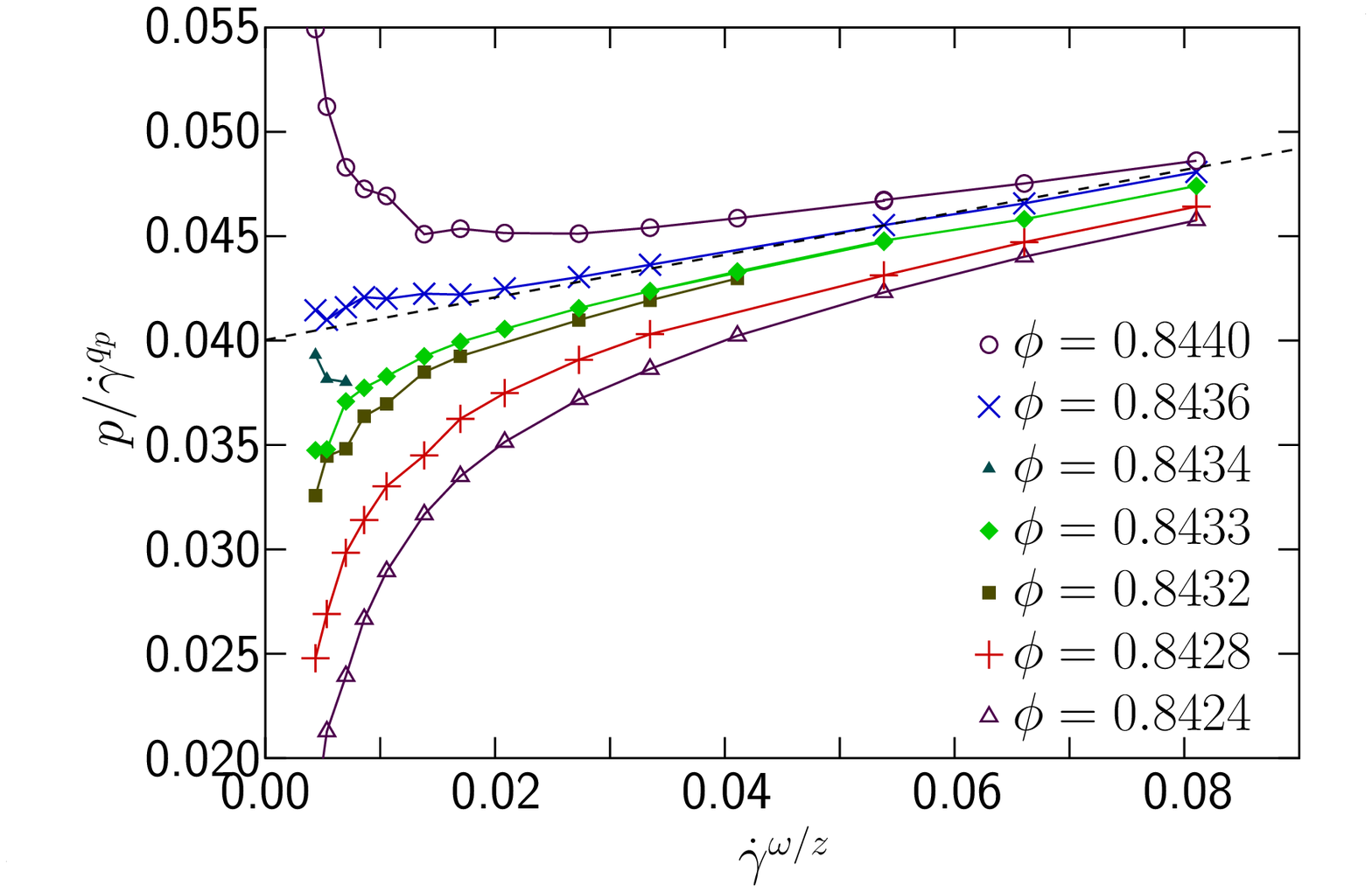}
  \includegraphics[width=4.1cm]{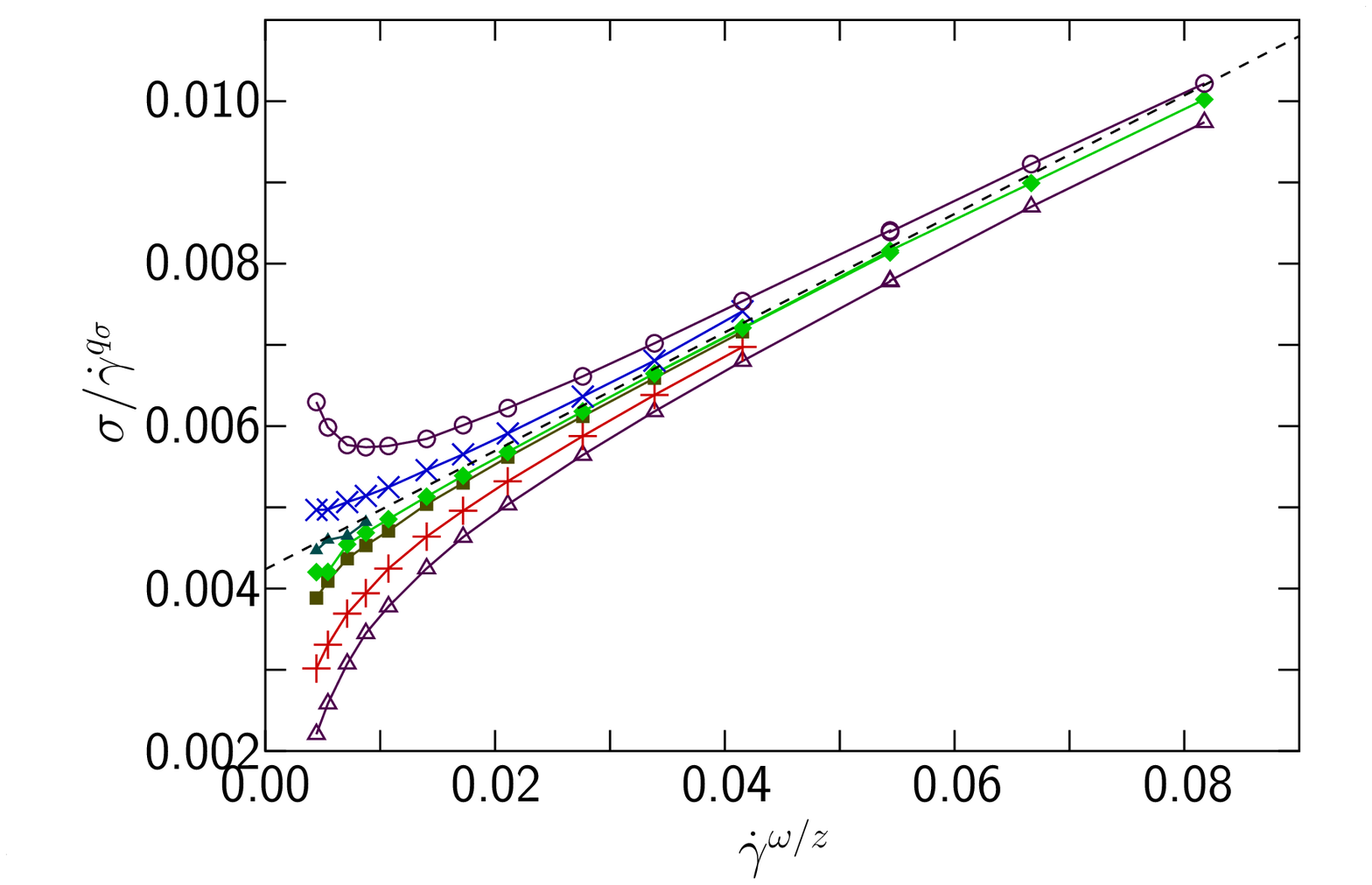}
  \caption{Illustration of results of the scaling analysis. The dashed lines are
    the behaviors at $\phi_J$ for $p$ and $\sigma$, respectively:
    $\O(\phi_J,\gdot)/\gdot^{q_\O} = g_\O(0) + \gdot^{\omega/z} h_\O(0)$.}
  \label{fig:O.q-gdot.w}
\end{figure}

The analyses of both pressure and shear stress work nicely when corrections to
scaling are included. A drawback with including the corrections is---beside the
more difficult analyses---that it is no longer possible to
determine $\phi_J$ directly from a simple plot as in \Fig{p-gdot}. The most
direct way to illustrate the determination of $\phi_J$ is shown in
\Fig{O.q-gdot.w} which displays $p/\gdot^{q_p}$ and $\sigma/\gdot^{q_\sigma}$
against $\gdot^{\omega/z}$, now with linear scales on both axes. Data at
$\phi_J$ should then fall on a straight line. Note the very different size of
the corrections, given by the slopes of the data.

\begin{figure}
  \includegraphics[width=6cm]{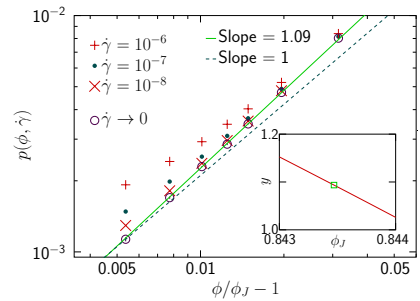}
  \caption{Alternative determination of the exponent $y_p$. The open circles are
    $p(\phi,\gdot\to0)$ from extrapolations of $p(\phi,\gdot)$. Assuming
    $\phi_J=0.84347$ the exponent becomes $y=1.09$, shown by the solid line. The
    dashed line corresponds to $y=1$. The inset shows how $y$ depends on the
    assumed $\phi_J$.}
  \label{fig:p-deltaphi}
\end{figure}

For $\phi$ well above $\phi_J$ the pressure decays algebraically in $\gdot$ and
this gives a means to determine the limiting value $p(\phi,\gdot\to0)$.  If we
can get reliable values, $p(\phi,\gdot\to0)$, at densities sufficiently close
above $\phi_J$ it should be possible to get another determination of $y_p$,
independent of the scaling analysis. \Fig{p-deltaphi} shows some of our
finite-$\gdot$ data together with such extrapolated values for densities down to
$\phi=0.848$. Fitting to the five points from $\phi=0.848$ through 0.856 (0.5\%
through 1.5\% above $\phi_J$) we find $y=1.09\pm0.04$ shown by the solid line,
in excellent agreement with $y=1.08$ from the scaling analysis. (The inset of
\Fig{p-deltaphi} shows how $y$ depends on the assumed $\phi_J$.) Similar
results, $y_p\approx 1.1$ have also been found before \cite{Majmudar_SLB,
  Heussinger_Chaudhuri_Barrat-Softmatter}.

The above results point to a good agreement between the exponent obtained from
the scaling analyses on the one hand, and the $\gdot\to0$ limit
of the pressure above $\phi_J$ on the other. This is entirely in accordance with
expectations from critical scaling.  This is in contrast to the claim in
\Ref{Tighe_WRvSvH} that the critical region is extremely narrow and doesn't
include densities away from $\phi_J$ in the limit $\gdot\to0$; the yield stress
is there taken to be governed by a different regime with a different
exponent, $y_\sigma=3/2$.

With the result $y\approx 1.1$ from two different analyses it becomes important
to try and reconcile this with the well established linear increase of the
pressure when marginally jammed packings are compressed above their respective
jamming densities \cite{OHern_Silbert_Liu_Nagel:2003,Chaudhuri_Berthier_Sastry}.
We speculate that the reason for this is that the ensemble of configurations
depends in a non-trivial way on $\phi$ in the vicinity of $\phi_J$, and that
this is so since \emph{the dynamic process that generates this ensemble} is
itself very sensitive to $\phi$.  It is then relevant to consider the behavior
in the quasistatic limit and to recall that the average time needed for the
minimization of energy in quasistatic simulations diverges as $\phi_J$ is
approached from above or from below.  (This parallels the more rapid jumping
between jammed and unjammed states reported in \Ref{Heussinger_Barrat:2009}.) A
dramatic change of the dynamical process that generates the ensemble suggests
that the ensemble itself would depend on $\phi$ in a non-trivial way.

To conclude, we have shown that pressure and shear stress from shearing
simulations are entirely consistent with the assumption of a critical behavior
when corrections to scaling are included in the analysis. We find
$\phi_J=0.84347\pm0.00020$ and that at $\phi_J$, both $p$ and $\sigma$  scale
as $\gdot^q$ with $q=0.28\pm0.02$.  In the limit $\gdot\to0$ both $p$ and $\sigma$ vanish as
$(\phi-\phi_J)^y$ with $y=1.08\pm0.03$.

This work was supported by Department of Energy Grant No.\ DE-FG02-06ER46298,
Swedish Research Council Grant No.\ 2007-5234, and a grant from the Swedish
National Infrastructure for Computing (SNIC) for computations at HPC2N.

%

\bibliographystyle{apsrev4-1}
\end{document}